\documentclass[10pt]{article}
\usepackage{graphicx}
\usepackage{amsmath}

\def\Title#1{\begin{center} {\Large #1 } \end{center}}
\def\Author#1{\begin{center}{ \sc #1} \end{center}}
\def\Address#1{\begin{center}{ \it #1} \end{center}}

\newcommand\pubblock{\rightline{\begin{tabular}{l} Proceedings of the Fifth Annual LHCP\\ \pubnumber\\
         \pubdate  \end{tabular}}}

\newenvironment{Abstract}{\begin{quotation} \begin{center} 
             \large ABSTRACT \end{center}\bigskip 
      \begin{center}\begin{large}}{\end{large}\end{center} \end{quotation}}

\newenvironment{Presented}{\begin{quotation} \begin{center} 
             PRESENTED AT\end{center}\bigskip 
      \begin{center}\begin{large}}{\end{large}\end{center} \end{quotation}}

\def\Acknowledgements{\bigskip  \bigskip \begin{center} \begin{large}
             \bf ACKNOWLEDGEMENTS \end{large}\end{center}}

\newcommand{\ctilde}{\tilde{c}}
\newcommand{\Arg}{\mathrm{Arg}}

\def\beq{\begin{equation}}
\def\eeq#1{\label{#1}\end{equation}}
\def\eeqn{\end{equation}}

\def\beqa{\begin{eqnarray}}
\def\eeqa#1{\label{#1}\end{eqnarray}}
\def\eeqan{\end{eqnarray}}

\let\bar=\overbar

\def\Dslash{\not{\hbox{\kern-4pt $D$}}}
\def\dslash{\not{\hbox{\kern-2pt $\del$}}}

\def\msb{{\bar{\ssstyle M \kern -1pt S}}}

\usepackage{color}

 \newcommand\pubnumber{ }

\newcommand\pubdate{\today}

\def\support{\footnote{Work supported by National Science Foundation of China under Grant No. 11405102 }}

\begin{document}

% large size for the first page
\large
\begin{titlepage}
\pubblock

%% Change the title, name, abstract
%% Title 
\vfill
\Title{  Probe CP violation in $H\to \gamma Z$ through forward-backward asymmetry  }
\vfill

%  if you need to add the support use this, fill the \support definition above. 
%   \Author{ FIRSTNAME LASTNAME \support }
\Author{ XIA WAN \support }
\Address{School of Physics $\&$ Information Technology, Shaanxi Normal University, Xi'an 710119, China}
\vfill
\begin{Abstract}

We suggest that the forward-backward asymmetry $(A_{FB})$ of the charged lepton in $gg\to H\to\gamma Z\to\gamma \ell^-\ell^+$ process could be used to probe the CP violating $H\gamma Z$ coupling when the interference from $gg\to\gamma Z\to\gamma \ell^-\ell^+$ process is included. With CP violation in $H\gamma Z$ coupling, the interference effect leads to a non-vanishing $A_{FB}$, which is also sensitive to the strong phase differences. The resonant and non-resonant strong phases together make $A_{FB}(\hat{s})$ change sign around Higgs mass $M_H$. For phenomenology study, we suggest the integral over one-side mass region below $M_H$ to magnify the $A_{FB}$ strength.

\end{Abstract}
\vfill

% DO NOT CHANGE 
\begin{Presented}
The Fifth Annual Conference\\
 on Large Hadron Collider Physics \\
Shanghai Jiao Tong University, Shanghai, China\\ 
May 15-20, 2017
\end{Presented}
\vfill
\end{titlepage}
\def\thefootnote{\fnsymbol{footnote}}
\setcounter{footnote}{0}
%

% normal size for the rest
\normalsize 

%% Your paper should be entered below. 

\section{Introduction}

To explain the observed matter-antimatter asymmetry in the universe,
some CP-violation sources beyond Standard Model (SM) are needed~\cite{Sakharov:1967dj}.
The Higgs boson discovered five years ago with mass around 125~GeV may provide clues
to study the source of CP violation.
Many papers have studied CP violation in Higgs couplings
such as $Ht\bar{t}$, $HZZ$, $HWW$ couplings~\cite{Ellis:2013yxa},\cite{Khachatryan:2014kca}.
In this work, we focus on $H\gamma Z$ coupling in the process $gg\to H \to\gamma Z\to \gamma \ell^- \ell^+$ at LHC.
Since there are only three final state momenta, 
the direct method to construct a CP violation observable fail. 
After considering the interference from a background process, there are some new CP violation observables: the forward-backward asymmetry ($A_{FB}$) of the leptons in $Z$ boson rest frame \cite{Chen:2014ona},\cite{Korchin:2014kha}, and the
 angle $\phi$ between the $Z$ production and decay planes
\cite{Farina:2015dua}. 
In this paper, we study the new CP violation observable $A_{FB}$ with the interference from the process $gg\to \gamma Z\to \gamma \ell^- \ell^+$
and discuss the its impact at current and future hadron colliders. 
This paper is a short report, a more detailed analysis could be found in Ref.~\cite{Chen:2017plj}.

\section{The effective model}
We use the following dimension-5 effective operators to describe the $gg \to H \to\gamma Z$ process,
\begin{equation}
        \label{eqn:HEF}
        \mathcal{L}_{\rm h} = \frac{c}{v}~h\,F_{\mu\nu} Z^{\mu\nu} + \frac{\ctilde}{2v}~h\,F_{\mu\nu}\tilde{Z}^{\mu\nu} + \frac{c_g}{v}~h\,G^a_{\mu\nu}G^{a\mu\nu}~,
\end{equation}
where $F$, $G^a$ denote the $\gamma$ and gluon field strengths, $a = 1,...,8$ are $SU(3)_c$ adjoint representation indices for the gluons, $v = 246$~GeV is the electroweak vacuum expectation value, the dual field strength is defined as $\tilde{X}^{\mu\nu}=\epsilon^{\mu\nu\sigma\rho}X_{\sigma\rho}$, $c$, $\ctilde$ and $c_g$ are complex numbers.

For simplicity, we require 
\beq
\Arg(c)=\Arg(\ctilde)~or~\Arg(c)=\Arg(-\ctilde)~.
\eeqn
After that it is convenient to define
\beq
\xi=tan^{-1}(\ctilde/c)~,
\label{eqn:xi}
\eeqn
where $\xi\in[0,2\pi)$. $\xi$ is a CP violation phase (weak phase) and we will show this when we discuss parity relation and CP transformation.

%%%%%%%%%%%%%%%%%%%%%%%%%%%%%%%%%%%%%%%%%%%%%%%%%%%%%%%%%%%%%%%%%%%%%%%%%
%%
%%   use this format to include an .eps figure into your paper
%%
\iffalse
\begin{figure}[htb]
\centering
\includegraphics[height=2in]{head_lhcp2017.jpg}
\caption{ Place the caption here}
\label{fig:figure1}
\end{figure}
%%%%%%%%%%%%%%%%%%%%%%%%%%%%%%%%%%%%%%%%%%%%%%%%%%%%%%%%%%%%%%%%%%%%%%%%%%%

See Figure \ref{fig:figure1} and Table \ref{tab:table1}. 

%%%%%%%%%%%%%%%%%%%%%%%%%%%%%%%%%%%%%%%%%%%%%%%%%%%%%%%%%%%%%%%%%%%%%%%%%
%%
%%   use this format to include a LaTeX table  into your paper
%%
\begin{table}[t]
\begin{center}
\begin{tabular}{l|ccc}  
Patient &  Initial level($\mu$g/cc) &  w. Magnet &  
w. Magnet and Sound \\ \hline
 Guglielmo B.  &   0.12     &     0.10      &     0.001  \\
 Ferrando di N. &  0.15     &     0.11      &  $< 0.0005$ \\ \hline
\end{tabular}
\caption{ place the caption here }
\label{tab:table1}
\end{center}
\end{table}
%%%%%%%%%%%%%%%%%%%%%%%%%%%%%%%%%%%%%%%%%%%%%%%%%%%%%%%%%%%%%%%%%%%%%%%%%%%
\fi

\section{Interference}
\begin{figure}[htb]
\centering
\includegraphics[width=0.5\linewidth]{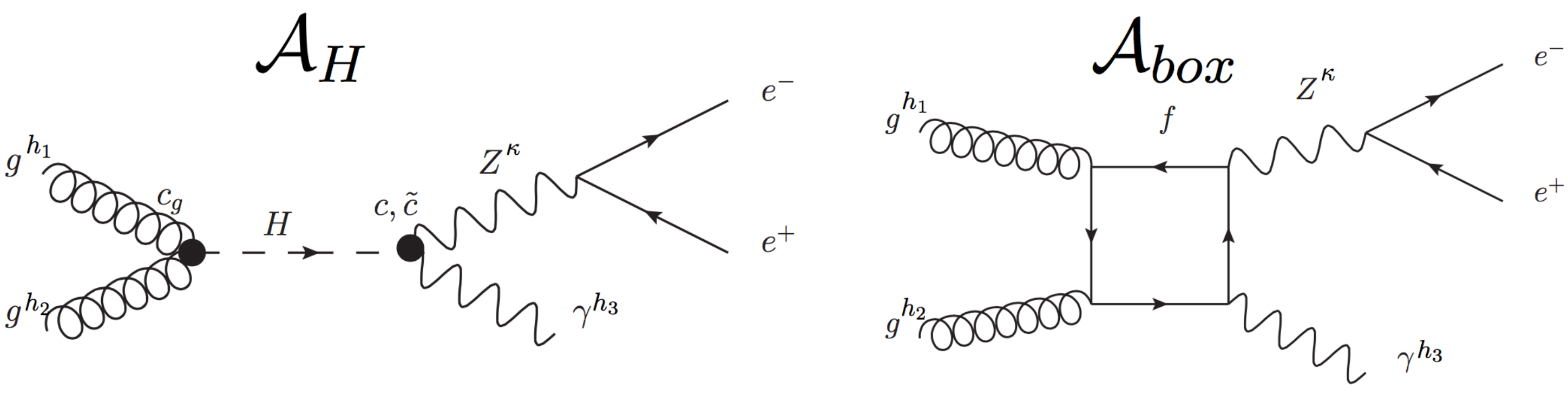}
\caption{\it The Feynman diagrams of $gg\to H\to\gamma Z\to\gamma \ell^-\ell^+$ and $gg\to\gamma Z\to\gamma \ell^-\ell^+$ processes. The amplitudes are noted as $\mathcal{A}_{H}$ and $\mathcal{A}_{box}$ respectively.}
\label{fig:gggammaz}
\end{figure}

The interference between the processes of $gg\to H\to\gamma Z\to\gamma \ell^-\ell^+$ and $gg\to\gamma Z\to\gamma \ell^-\ell^+$ are considered. Their Feynman diagrams are shown in Fig.~\ref{fig:gggammaz}. 
The intermediate $Z$ boson is considered to be on-shell with a narrow-width approximation. After that the 
 $2\to 3$ process could be factorized into a $2\to 2$ process times a $1\to 2$ process and the total squared amplitude is
\vspace{-0.5em}
\beq
\big|\mathcal{A}|^2
=\sum_{h_i}\left|\sum_{\kappa=+,0,-}
[\mathcal{
A}_H^{2\to2}+\mathcal{
A}_{box}^{2\to2}]^{h_1 h_2}_{h_3 \kappa}
[\mathcal{
A}^{1\to2}]^{-\kappa}_{h_4 h_5}\right|^2~,
~[\mathcal{A}_H^{2\to2}]^{h_1 h_2}_{h_3 \kappa}=
[\mathcal{A}_H^{SM~2\to2}]^{h_1 h_2}_{h_3 \kappa}
\times e^{-i\kappa\xi}~,
\eeqn
where $h_i$s are the helicities of gluons and photon, $\kappa$ is the helicity of $Z$ boson.

$[\mathcal{A}_{H}^{2\to2}]^{h_1 h_2}_{h_3 \kappa}$ has parity relations as
\beq
[\mathcal{A}^{2\to2}_H]^{-h_1 -h_2}_{-h_3 -\kappa} =
[\mathcal{A}^{2\to2}_H]^{h_1 h_2}_{h_3 \kappa}
\bigg|_{\xi\leftrightarrow-\xi}~,
\eeqn
which shows $\xi$ changes sign under CP transformation and thus is a weak phase. 

\section{Kinematics and the Source of $A_{FB}$}

\begin{figure}[htb]
\centering
\includegraphics[width=0.25\textwidth]{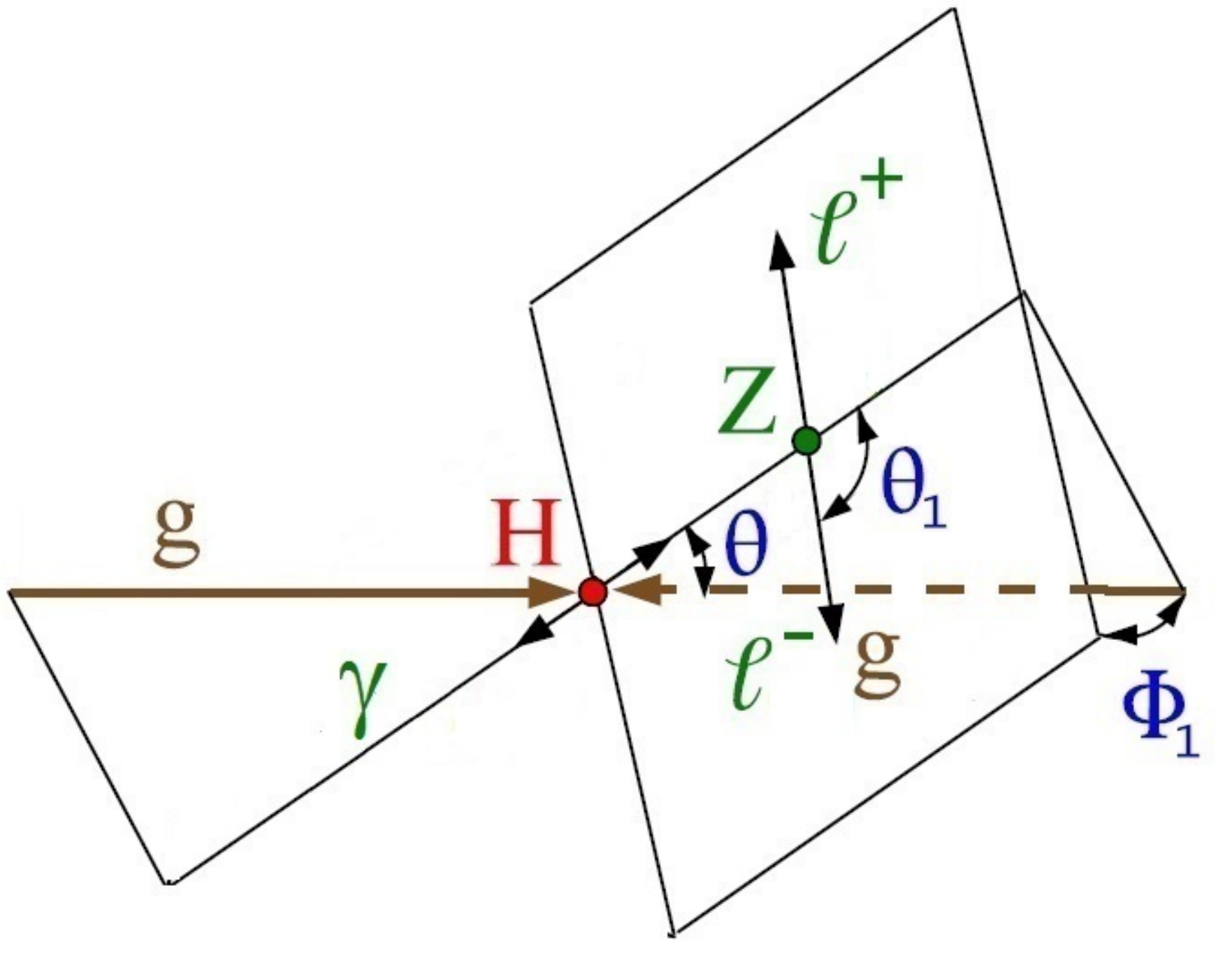}
\caption{\it The kinematic angles for $gg\to H\to\gamma Z \to \gamma \ell^-\ell^+ $ process.
$\theta$ is the polar angle of $Z$ boson in $H$ (or $gg$) rest frame.
$\theta_1$ is the angle of $\ell^-$ in $Z$ boson rest frame.
The z-axis of $Z$ boson rest frame is defined as the $Z$ boson production momentum direction in $H$ rest frame. $\phi_1$ is the angle between $Z$ boson production and decay planes.  }
\label{fig:hzgammakinematics}
\end{figure}

We only
need five variables to character the full kinematics.
The independent variables are the two squared invariant masses $\hat{s}$ and $s_{45}$, and the three angles $\theta$, $\theta_1$ and $\phi_1$.
Fig.~\ref{fig:hzgammakinematics} illustrates the three angles.

The forward-backward asymmetry ($A_{FB}$) in proton-proton collision is
\vspace{-0.5em}
\beqa
A_{FB}
&\equiv&\frac{N_F-N_B}{N_F+N_B}
=\frac{(\int^1_0-\int^0_{-1})d\cos{\theta_1}\int_I d\sqrt{\hat{s}}
\sqrt{\hat{s}}G(\hat{s})
\frac{d\hat\sigma(\hat{s},\theta_1)}{d(\cos\theta_1)}}
{(\int^1_{-1})d\cos{\theta_1}\int_I d\sqrt{\hat{s}}
\sqrt{\hat{s}}G(\hat{s})
\frac{d\hat\sigma(\hat{s},\theta_1)}{d(\cos\theta_1)}}~\\
&\propto& \int_I d\sqrt{\hat{s}}
\sqrt{\hat{s}}G(\hat{s})
\operatorname{Im}[\tilde{\sigma}_{H,box}^{2\to2}]_{++}\sin\xi,
\label{eqn:sinxi}
\eeqan
\vspace{-1em}
\beqa
\operatorname{Im}[\tilde{\sigma}_{H,box}^{2\to2}]_{++}
&=&\operatorname{Im}\sum_{h_1,h_2,h_3}[\mathcal{
A}_H^{2\to2}]^{h_1h_2}_{h_3 +}(\hat{s},\theta)[\mathcal{
A}_{box}^{\ast2\to2}
]^{h_1h_2}_{h_3 +}(\hat{s},\theta)~,
\eeqan
where $\sqrt{\hat{s}}=M_{\gamma Z}$, $s$ being the total hadronic c.m. energy and $G(\hat{s})$ is the gluon-gluon luminosity function, 
$\int_I$ represents an mass region to be integrated.
Because of several non-zero strong phases, the integrand $\operatorname{Im}[\tilde{\sigma}_{H,box}^{2\to2}]_{++}$
changes sign around resonance peak as shown in the simulation.

\section{Simulation}

\begin{figure}[htb]
\centering
\includegraphics[width=0.3\linewidth]{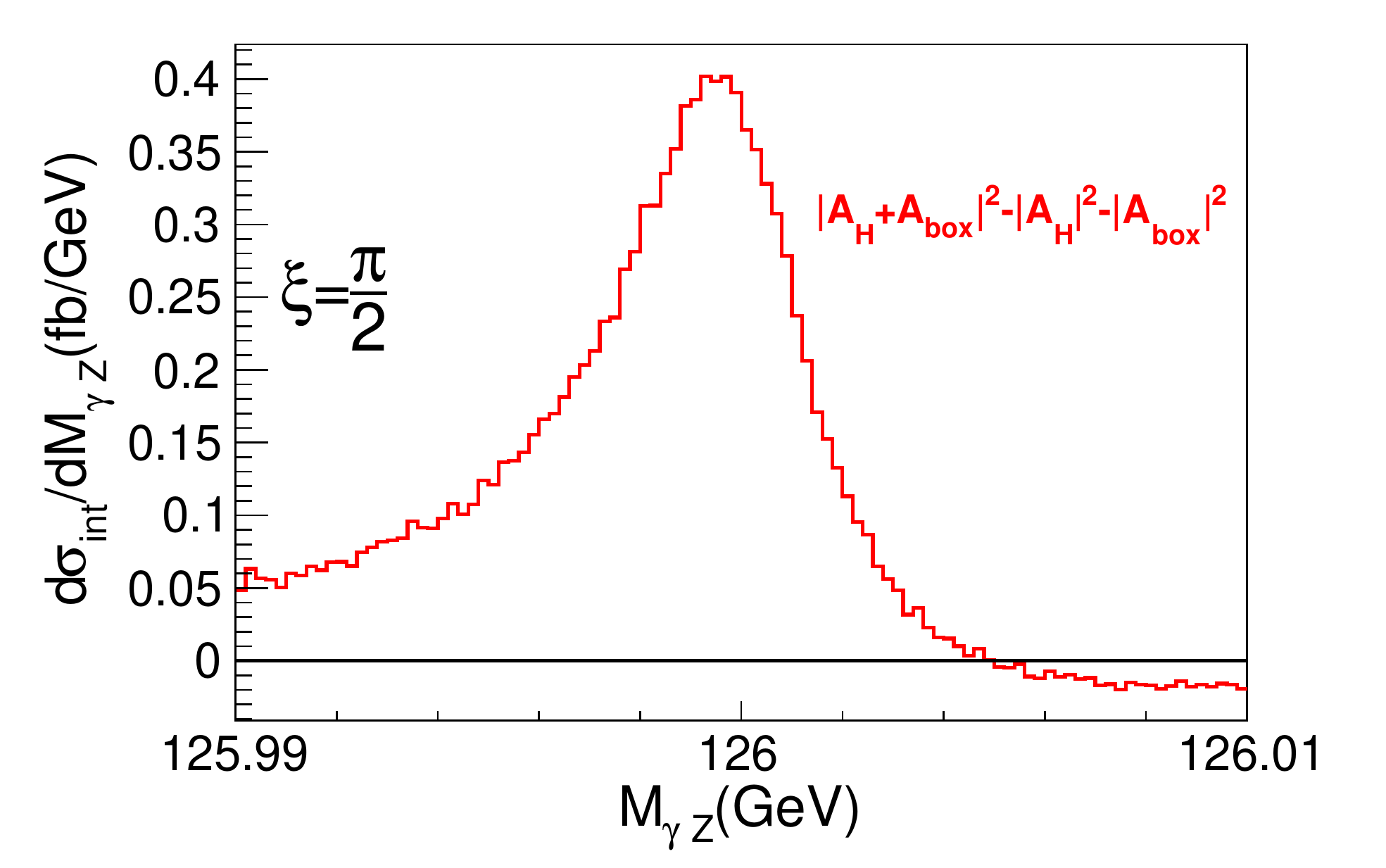}
\includegraphics[width=0.3\linewidth]{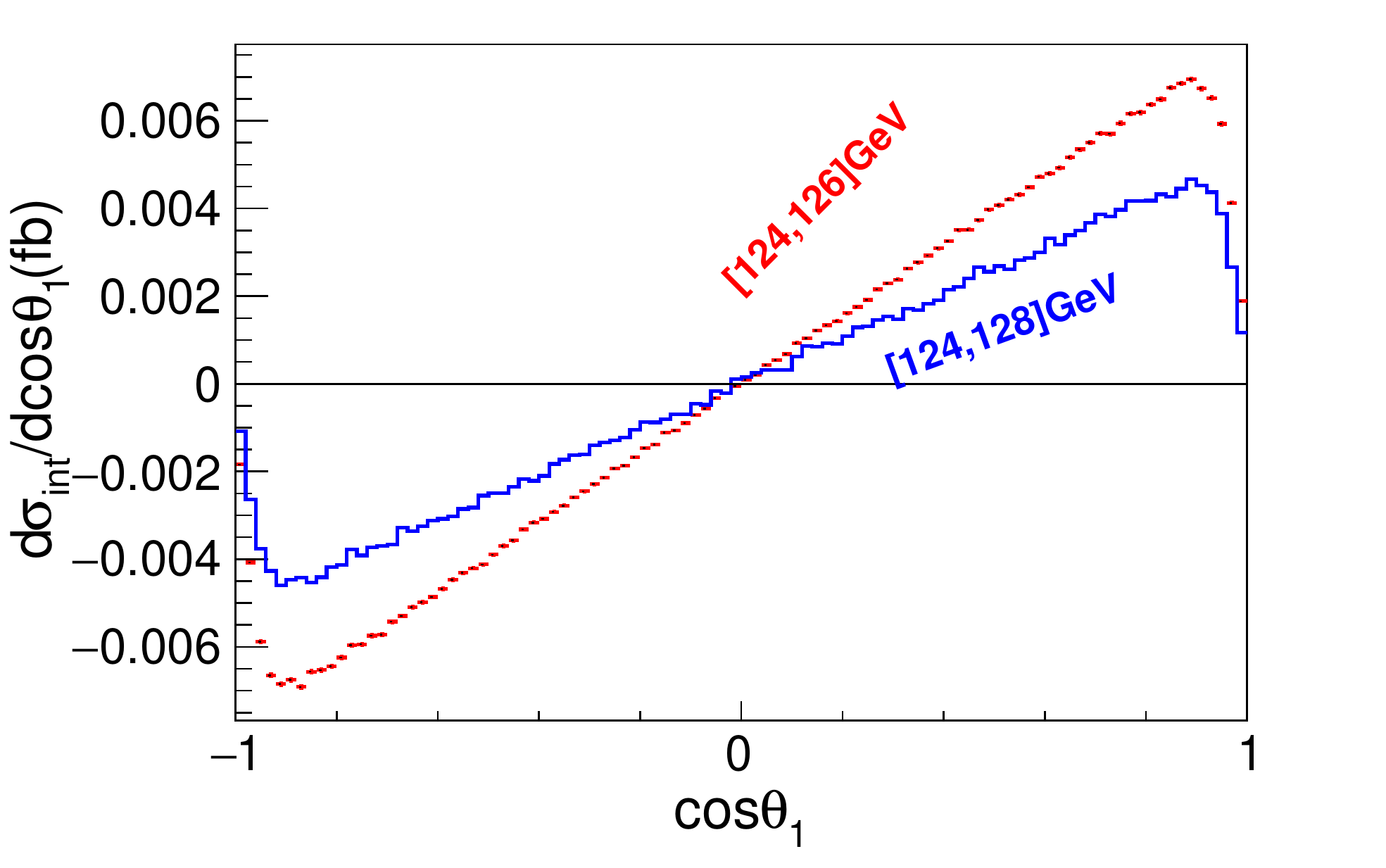}
\caption{\it Left panel: The $M_{\gamma Z}$ differential cross section of
the interference part, which changes sign around the resonance peak. Right panel: $d\sigma_{int}/d\cos{\theta_1}$ versus $\cos{\theta_1}$ for $[124,126]$~GeV and $[124,128]$~GeV integral region. The slope is equal to the numerator value of $A_{FB}$.}
\label{fig:afb}
\end{figure}

Based on a modified \texttt{MCFM} package, the simulation are generated for a proton-proton collider with $\sqrt s=14$~TeV. The selection criteria include: $p^{\gamma}_T>20$~GeV, $|\eta^{\gamma}|<2.5$ and $M_{\ell^-\ell^+}\in$ [66, 116]~GeV.
In Figure~\ref{fig:afb},
the left panel shows the integrand of $A_{FB}$ numerator changes sign 
around the resonance peak, the right panel shows the slope (also $A_{FB}$ numerator) is larger when integral region is half of the resonance region.
 Table~.\ref{tab:table1} shows $A_{FB}$  
could be enhanced if integrate over the half resonance region. 

Experimentally the invariant mass resolution could dilute the asymmetric component of interference and minify the $A_{FB}$ value. The resonance mass uncertainty may
make it difficult to get a half resonance region and thus the $A_{FB}$ value would be also deviate to the theoretical prediction. Nevertheless, it is still better to consider the integral over one side of the resonance peak and the $A_{FB}$ value would still be larger than if integrated over the whole resonance region.  

\begin{table}[t]
\begin{center}
\begin{tabular}{c|c}
\hline
Integral mass region (GeV) & $A_{FB}$ \\  %\tabularnewline
 \hline
 [124, 126] &  $0.008/1.4\sim0.57\%$  \\  %\tabularnewline
 \hline
[124, 128]  &  $0.005/2.8\sim0.18\%$   \\ %\tabularnewline
\hline
\end{tabular}
\caption{\it $A_{FB}$ values when integrating over half and whole resonance regions.}
\label{tab:table1}
\end{center}
\end{table}

\section{Conclusion and discussion}
In this work we construct a model with general CP violation phase $\xi$ from $H\gamma Z$ coupling. By calculating the interference effect between $gg\to H\to\gamma Z\to\gamma \ell^-\ell^+$ and $gg\to\gamma Z\to\gamma \ell^-\ell^+$ processes, we confirm that the forward-backward asymmetry $A_{FB}$ of charged leptons in the $Z$ rest frame is a CP-violation observable, and is proportional to $\sin\xi$. By studying the lineshape of the integrand, we propose to do integral of $M_{\gamma Z}$ over half of the resonant mass region to enhance $A_{FB}$. After detailed simulations using modified \texttt{MCFM}, we estimate the $A_{FB}$ could reach about $0.6\%$.
Even though, the significance is relatively small and hard to be observed at the HL-LHC.
%%  if necessary
\Acknowledgements
I am grateful to Xuan Chen, Gang Li and Youkai Wang for fruitful discussions.

%\bibliographystyle{apsrev}
%\bibliography{reference}

\end{document}